# Epitaxial Oxide Thin Films with High Entropy and Anomalous Crystallinity

Joseph Petruska[1], Christina M. Rost[2], Jon-Paul Maria[1], and *Saeed S. I. Almishal[1]

[1]*Department of Materials Science and Engineering, The Pennsylvania State University, University Park, PA 16802, USA*
[2]*Department of Materials Science and Engineering, Virginia Polytechnic Institute and State University, Blacksburg, VA 24061, USA*

**Corresponding Authors:** Saeed S. I. Almishal ssa5409@psu.edu

**Keywords:** High-entropy oxides, epitaxy, thin film growth, pulsed laser deposition, anomalous crystallinity

## Abstract

Pulsed laser deposition (PLD) expands the compositional and structural possibilities of high-entropy oxide (HEO) thin films through far-from-equilibrium growth and kinetic arrest of metastable configurations. This review connects growth kinetics and epitaxial constraints to the roles of individual cations in shaping the resulting films. We focus primarily on rock salt MgCoNiCuZnO and its derivatives, with selected fluorite, spinel, and perovskite examples illustrating broader applicability. In MgCoNiCuZnO, Co valence and Cu coordination illustrate how chemistry and growth conditions jointly enable lattice tuning and coherent nanostructure formation, while aliovalent and size-misfit additions expand the accessible defect and compositional landscapes. These coupled degrees of freedom underpin anomalous crystallinity, the coexistence of extensive chemical disorder with exceptional crystalline quality and epitaxial order, including full coherence in selected systems, particularly rock salts grown on MgO. This combination opens opportunities to discover and engineer functional configurations in high-quality single-layer films and heterostructures.

## 1 | Introduction and Scope

Entropy engineering expands the possibilities for oxide materials design by combining chemical diversity within a shared crystalline framework. In 2015, entropy stabilization was demonstrated in equimolar MgCoNiCuZnO, establishing configurational disorder as a means of stabilizing solid solutions whose constituents otherwise favor competing structures.

Numerous reviews have examined the thermodynamics, synthesis, structures, and functional properties of high-entropy oxides (HEOs).[1–8] The present review provides a dedicated discussion of far-from-equilibrium thin-film growth enabled by pulsed laser deposition (PLD). Its central focus is how growth conditions and cation chemistry together determine which structural, valence, and defect configurations are accessed and retained. PLD combines energetic deposition onto comparatively cool substrates with controllable growth parameters that govern cation incorporation, relaxation towards equilibrium, and kinetic arrest.[1,9–11] In HEO films, these controls can extend compositional accessibility and tune lattice parameters and local ordering while preserving high crystalline quality and, under suitable conditions, epitaxial coherence.[1,9–14] This coexistence of substantial chemical disorder with high crystalline quality and epitaxy is termed anomalous crystallinity.[10] It provides rich opportunities for coherent heterostructure design and untapped property engineering, as explored in this review.

We focus primarily on the rock salt MgCoNiCuZnO system, referred to here as J14, its four-component derivatives, and compositions containing additional aliovalent or size-misfit cations. Selected fluorite, spinel and perovskite systems illustrate how these considerations extend to other oxide families. **Section 2** discusses growth parameters and kinetic arrest; **Section 3** examines individual cation roles, including Co valence, Cu coordination, and sixth-cation additions. **Section 4** develops the concept of anomalous crystallinity and its implications for heterostructure design, followed by a brief discussion on the resulting functional properties in **Section 5.**

## 2 | Kinetics and Growth Parameters

### 2.1 | PLD Kinetics and Extended Solubility

Owing to their high configurational entropy, HEOs exhibit complex energy landscapes containing numerous metastable macrostates and microstates. Careful control of synthesis conditions can tune the balance among entropic stabilization, enthalpic driving forces, and kinetic barriers, thereby determining which states are accessed and subsequently retained upon quenching.[1,4,11,15,16] The qualitative time-temperature-transformation diagram in Figure 1(a) compares the kinetic pathways of common HEO synthesis methods, including solid-state synthesis, solution-based synthesis, and physical vapor deposition (PVD), showing how their distinct thermal histories influence phase formation and kinetic arrest.[1] Among the various synthesis techniques, pulsed laser deposition (PLD), a PVD technique, is unequivocally well suited to navigate HEO energy landscapes through (1) far-from-equilibrium pathways as detailed below, and (2) independently tunable synthesis conditions (as detailed in Section 2.3 and 2.4). PLD can thereby access atomic and electronic configurations, and associated physical properties, beyond those realized through near-equilibrium synthesis methods.[9,17,18]

In PLD, laser pulses lasting tens of nanoseconds ablate a target and generate a high energy transient plume ($10^4$-$10^5$ K effective temperatures) whose species reach the comparatively cool substrate on microsecond time scales and rapidly condense to form a thin film. The kinetic energies of the arriving species, together with substrate structure, temperature, ambient pressure, and pulse repetition rate, influence this rapid condensation, mixing, nucleation, and surface diffusion.[9–12,18] These coupled controls can incorporate cations that are difficult to accommodate through near-equilibrium bulk synthesis.[9,10,14,17,18] Between pulses (0.05-1 s at repetition rates of 1-20 Hz), deposited species can diffuse across the surface, while continued deposition progressively buries them.[11] When cation redistribution and phase separation cannot keep pace with growth and subsequent cooling, the film can retain metastable states, including high-symmetry chemically disordered solid solutions, with distinct structural, valence, and defect configurations.[1,9–12,14,16,18,19] Figure 1(b) illustrates how varying these growth conditions changes the deposition pathway and creates opportunities to retain metastable states.[1]

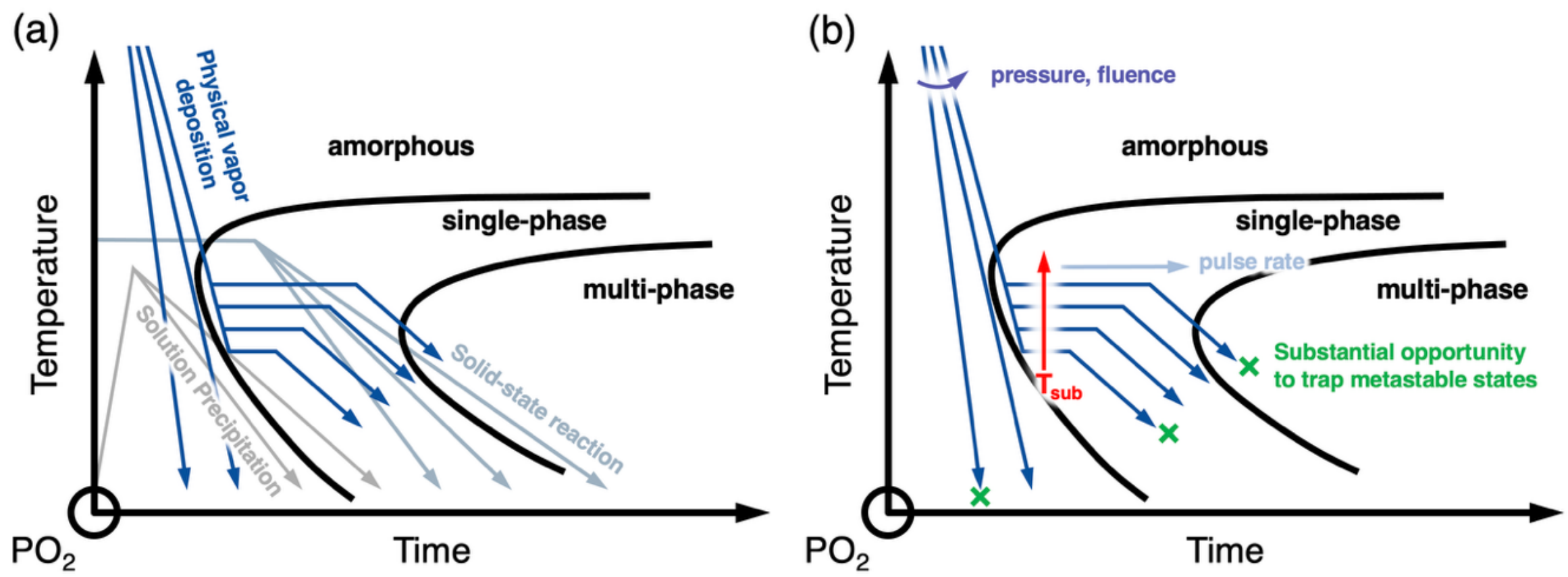

**Figure 1:** (a) Qualitative time-temperature-transformation (TTT) diagram illustrating the different kinetics of common high-entropy oxide (HEO) synthesis methods. Cool blue indicates solid-state reactive sintering, gray indicates solution combustion or powder precipitation techniques, and blue indicates high-energy physical vapor deposition techniques such as pulsed laser deposition (PLD) and sputtering. Note the variable cooling rates possible during reactive sintering. During vapor deposition, choice of growth temperature and growth time results in many possible pathways through the diagram highlighted in (b). Varying growth conditions for physical vapor deposition provides a substantial opportunity to trap metastable states. Figure 1 is adapted and modified from Kotsonis, G. N., Almishal S.S.I, *et al.* High-entropy oxides: Harnessing crystalline disorder for emergent functionality. *J. Am. Ceram. Soc.* **106**, 5587–5611 (2023). [Ref 1]

The prototypical rock salt HEO MgCoNiCuZnO (J14) exemplifies the extended solubility enabled by PLD. Adding $Sc^{3+}$ or $Cr^{3+}$ as an equimolar sixth cation challenges the charge balance of the predominantly divalent host lattice, yet suitable PLD conditions produce coherent highly crystalline rock salt films containing either cation[10]. Their contrasting ionic sizes and associated defect chemistry, combined with substrate-temperature control, enable more than 7% variation in the out-of-plane lattice parameter between the two 6-component compositions (ScMgCoNiCuZnO (JSc) and CrMgCoNiCuZnO (JCr)).[10] $Ca^{2+}$ tests a different limit: although it is isovalent and favors rock salt coordination, its ionic radius is approximately 38% larger than the average J14 cation radius.[14] Bulk synthesis incorporates at most approximately 4% Ca into Cu-containing J14, whereas PLD extends incorporation to approximately 10%.[14] Removing Cu from the composition further permits an equimolar 20% Ca CaMgNiCoZnO film.[14,20] These examples show that PLD expands the accessible compositional landscape, while the limits of incorporation still depend on the added cation's size, valence, and local bonding.[9,10,14] We examine these cases further in Sections 3.4 and 4.1.

## 2.2 | Substrates and Epitaxy

In thin-film growth, texture describes a preferred crystallographic orientation among the grains of a film, whereas epitaxy describes a defined crystallographic orientation relationship between the film and its substrate. Coherent epitaxy further requires lattice registry across the interface without misfit dislocations.[9,21] When the film also maintains the substrate's in-plane lattice spacing throughout its thickness, its growth is described as pseudomorphic.[10,11,14] For J14-derived films on MgO, the terms coherent, commensurate, and pseudomorphic are often used interchangeably to describe this high-crystalline-quality lattice matched state.[9–12,14,15]

The role of epitaxy in J14-derived films formation and stability depends on composition and synthesis conditions: it can improve crystalline quality particularly when coherency is achieved and, in some cases, influence competition between rock salt and secondary phases during growth.[9,10,15] JSc illustrates the first role. Under suitable PLD conditions, coherent epitaxial film on MgO, textured film on polycrystalline alumina, and non-epitaxial film on amorphous silica exhibit single-phase rock salt diffraction without detectable $Sc_2O_3$ secondary phase.[9] Increasing background pressure promotes phase separation on both MgO and alumina, indicating that deposition conditions play a stronger role than the type of epitaxy in suppressing the secondary $Sc_2O_3$ in the JSc films.[9] Epitaxial coherent growth on MgO nevertheless produces substantially sharper diffraction features, indicating higher crystalline quality.[9] For JCr, the substrate comparison suggests a role in phase evolution. At 500 °C, films on MgO remain predominantly rock salt by XRD, although peak asymmetry and electron diffraction reveal local spinel-like ordering. Films grown at the exact same conditions on c-plane sapphire instead exhibit distinct spinel reflections.[10] The Mg-free J14-derived composition presents a closely similar case: at 400 °C substrate temperature, the growth on MgO diffraction pattern suggests a possible spinel contribution, which is clearly resolved in the corresponding sapphire film.[15] Together, these cases suggest that the epitaxial and coherent growth on MgO can retain the rock salt matrix and limit the competing phases, even when it does not fully suppress local ordering.

## 2.3 | Substrate Temperature

Substrate temperature plays two related roles during PLD growth: (1) it governs whether the initially condensed rock salt solid solution is kinetically retained during growth and subsequent cooling, and (2) within the single-phase growth window, it controls how far the film's structural, valence, and defect configurations evolve toward equilibrium.

The first role is evident when limited atomic mobility helps retain a metastable phase. For example, at lower temperatures, reduced surface and subsurface mobility can kinetically arrest configurations that are further from equilibrium that would otherwise rearrange or phase separate

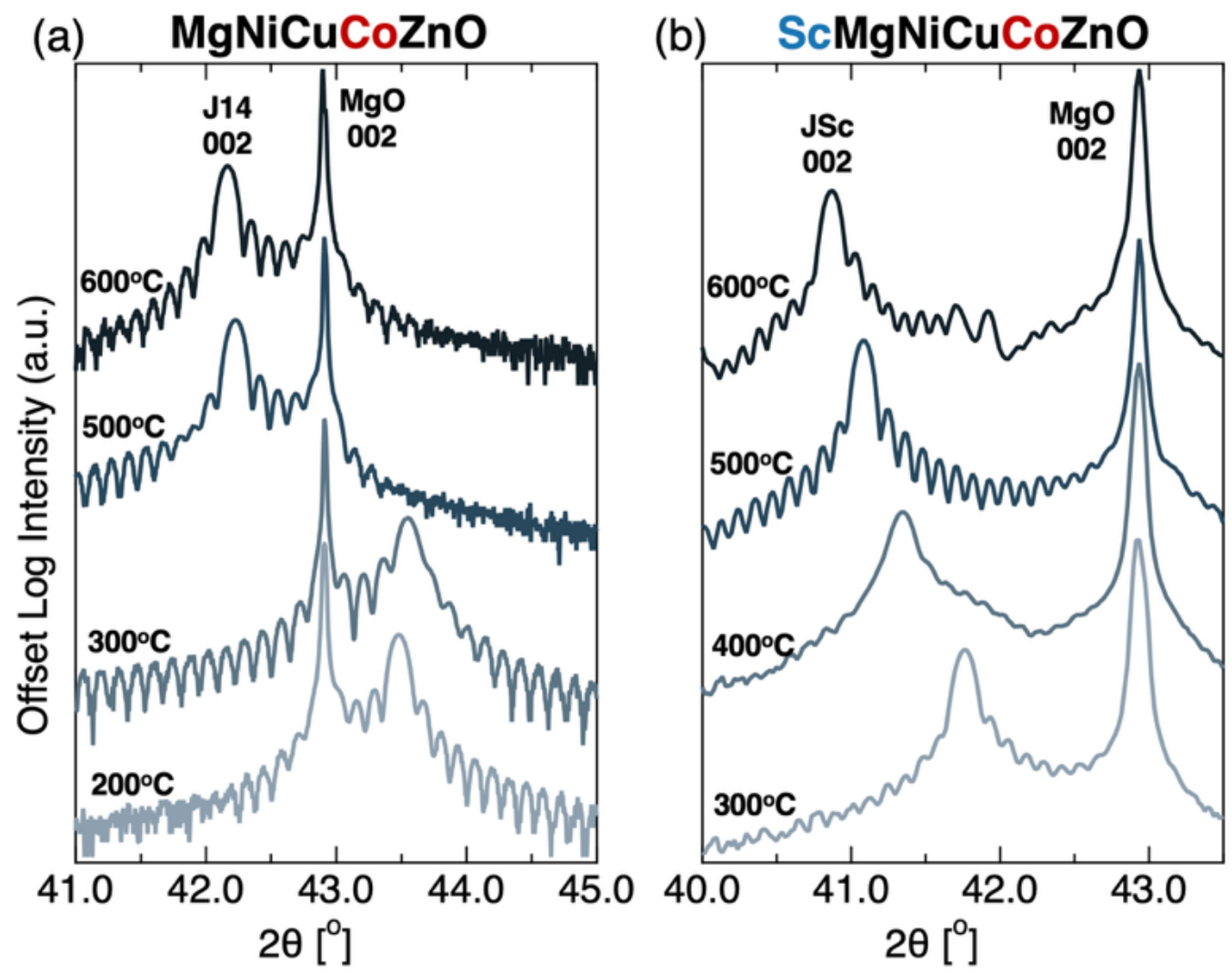

**Figure 2: Temperature-dependent out-of-plane lattice parameter change in rock salt high-entropy oxide (HEO) thin films.** $\theta$–$2\theta$ X-ray diffraction scans around the (002) reflection for two rock salt oxide compositions grown on MgO(001) under 50 mTorr flowing $O_2$: (a) the prototypical five-component HEO MgCoNiCuZnO (J14), and (b) the Sc-containing six-component HEO ScMgCoNiCuZnO (JSc), deposited at different substrate temperatures as indicated. Figure 2 is adapted from Almishal, S. S. I. *et al.* Anomalous Crystallinity and Magnetism in Chemically Disordered Coherent Heterostructures. *Adv. Mater.* **38**, e73998 (2026). [Ref 10]

during warmer growth.[10,14,15] Ca-containing J14-derived films illustrate this sensitivity: at fixed laser fluence, growth at or above 500 °C produces an additional X-ray diffraction (XRD) feature consistent with a possible Ca-rich, strain-relaxed component, absent at lower growth temperatures. Substrate temperature therefore constrains the growth window for high-quality, single-phase Ca-containing films.[14] As discussed in the following section, laser fluence and substrate temperature together influence which metastable states form and how far the resulting film remains from equilibrium.

The second role is evident in how the substrate temperature can tune valence and defect states within a single-phase growth window. In J14 films on MgO, for example, raising the growth temperature from 350 to 375 °C expands the out-of-plane lattice parameter by approximately 3% while preserving coherence, Figure 2(a).[10] This abrupt expansion reflects a shift from $Co^{3+}$ toward the larger $Co^{2+}$ ion coupled with $Cu^{2+}$ Jahn-Teller distortions, as detailed in sections 3.2 and 3.3.[12,13,15] JSc responds more gradually as depicted in Figure 2(b): its out-of-plane lattice parameter increases by approximately 2% from 300 to 600 °C, while pronounced Pendellösung fringes persist even as it reaches 4.41 Å, 4.7% larger than MgO.[10] These temperature-driven changes demonstrate substantial lattice and valence tunability while maintaining high crystalline quality available to these epitaxial high entropy oxide thin films.

## 2.4 | Other PLD Growth Parameters: Laser Fluence, Pressure, and Pulse Rate

Beyond substrate temperature, laser fluence, background pressure and gas composition, and pulse repetition rate provide coupled controls over PLD growth. Fluence influences the energy and amount of material ejected from the target, while collisions with the background gas modify the species' kinetic energies before arrival.[22–24] Gas composition also influences oxidation conditions at the growth surface.[25] Repetition rate controls the interval between deposition events, while total pulse count influences film thickness.[11] Together, these parameters govern phase retention, crystalline quality, and local ordering within the deposited film.[9,10,14,16,18]

Ca-containing J14-derived films illustrate the composition-dependent influence of laser fluence. At a substrate temperature of 400 °C, films grown at fluences of 0.75–1.5 J/cm$^2$ exhibit lower crystalline quality, secondary phases, and additional orientations. Increasing the fluence to 1.87 J/cm$^2$ improves crystalline quality and suppresses the additional orientations, although further increases do not continually improve film quality.[14] By comparison, J14 grows with high crystalline quality at 1.2 J/cm$^2$ and, in our experience, at lower fluences.[12] Thus, under the conditions examined, Ca incorporation shifts the fluence window for high-quality growth toward higher values than those suitable for the parent composition.[14] Beyond rock salts, fluence and substrate temperature can jointly stabilize high symmetry phases not attained in bulk synthesis and tune ordering on the anion sublattice. For example, the lanthanide HEO $Ce_{0.2}(Y,La,Pr,Sm)_{0.8}O_{2-\delta}$ adopts bixbyite under near-equilibrium bulk processing[16]. Films grown at 1.5 J/cm$^2$ and 650 °C exhibit the bixbyite (006) reflection, whereas growth at 2.9 J/cm$^2$ and 450 °C suppresses this reflection, producing a globally fluorite-like structure with weak residual ordering detected by electron diffraction.[16] These observations are consistent with higher-fluence deposition combined with lower substrate temperature favoring retention of a higher-symmetry structure, and in this case, with a more disordered oxygen-vacancy framework.[16,19]

Background pressure and gas composition further influence phase retention through plume thermalization and oxidation conditions.[9] Increasing pressure generally increases gas-phase collisions and reduces the kinetic energy of arriving species, potentially limiting retention of phases that depend on energetic deposition.[9] For JSc, lower pressures favor rock salt retention, whereas higher pressures promote phase separation on both MgO and alumina.[9,10] Changing the $O_2$/Ar ratio modifies both plume scattering and the oxidizing environment.[9] Pressure and gas composition must therefore be considered together when establishing a growth window, particularly for compositions containing valence-sensitive cations.[9,10,12] Indeed, the appropriate oxygen activity depends on the oxidation states required by the desired phase relative to its competing phases. While oxygen-containing atmospheres favor growth of the rock salt and fluorite HEOs discussed above, some perovskites require oxygen-lean conditions.[9,10,12,16,18] In $Sr_{0.95}(Ti,Cr,Nb,Mo,W)O_3$, the perovskite requires an average B-site valence near 4+, whereas competing phases present in the bulk ceramic target, $SrCrO_4$, $Sr_2Nb_2O_7$, $SrMoO_4$, and $SrWO_4$, contain $Cr^{6+}$, $Nb^{5+}$, $Mo^{6+}$, and $W^{6+}$, respectively.[18] Epitaxial perovskite films are grown by PLD at 850 °C in 50 mTorr Ar to limit oxidation and suppress these competing phases.[18] Oxygen-lean growth can therefore favor retention of a high-entropy phase with lower cation oxidation states than those of competing phases formed during near-equilibrium synthesis.[18] More details on this system are in Section 4.5.

Pulse repetition rate and film thickness provide additional controls over relaxation towards equilibrium.[11] For J14 grown at 400 °C and 1.2 J/cm$^2$, reducing the repetition rate from 5 to 1 Hz favors coherent CuO-rich nanotweed formation within the global rock salt phase, whereas increasing the pulse count from 1200 to 6000 favors spinel-structured nanocuboids formation.[11,26] Combining slower deposition with greater thickness produces both features. All four films remain pseudomorphic to MgO. The nanoscale ordering is resolved by electron microscopy and diffraction, although the nanocuboids are not directly identified as a separate spinel phase by XRD. These results indicate that inter-pulse relaxation and thickness-dependent strain energy can tune nanoscale configurations within a coherent matrix.[11] The associated Co and Cu chemistry is discussed in Sections 3.2 and 3.3.

Together, these growth parameters define composition-dependent windows for retaining high-entropy phases and accessing local chemical or structural ordering within them. The resulting film is described by both its overall phase and its retained microstate, characterized by cation distribution, valence, strain, and nanoscale order. Growth conditions govern which microstates are sampled and which are frozen in; composition defines the ensemble of microstates available to be sampled.[9–11,14,16,18,19] The next section examines how individual cations shape this ensemble and the properties that emerge from the microstate retained by kinetic arrest.

## 3 | Cation Roles and Stoichiometry

In HEOs, individual cations play distinct roles in phase evolution, stability, and properties.[10,15] Section 3.1 examines these roles, followed by dedicated discussions of Co and Cu in Sections 3.2 and 3.3, respectively, emphasizing their influence on lattice parameters and nanostructure formation. Finally, Section 3.4 considers sixth-cation additions to J14, examining how they expand the accessible structural, valence, and defect landscapes, and the growth and chemical strategies used to accommodate them.

### 3.1 | Untangling Individual Cation Roles in MgCoNiCuZnO

The roles of individual cations in J14 can be examined through its five equimolar four-component derivatives, each formed by removing Mg, Co, Ni, Cu, or Zn.[15] Bulk synthesis above 875 °C provides a reference: Mg, Ni, and Co favor rock salt formation, whereas Cu and Zn favor competing structures.[15] Under PLD conditions, lower growth temperatures and kinetic arrest modify these roles, particularly for Co. Across substrate temperatures of 200-600 °C, removing any cation except Cu narrows the single-phase growth window relative to J14.[15] Reciprocal space maps around MgO (204) show that the selected single-phase films retain the substrate's in-plane lattice spacing despite distinct out-of-plane responses, Figure 3(a–e).[15] Ni removal in Figure 3(c) restricts single-phase growth to 300 °C, identifying Ni as the strongest stabilizer, followed by Mg. Cu removal preserves the broadest growth window indicating a stronger destabilizing influence than Zn, , Figure 3(b and d).[15] Figure 3(a) shows that Co removal also permits a relatively broad window compared with the other derivatives, indicating that Co can hinder homogeneous rock salt formation under these conditions despite stabilizing it in bulk.[15] Additionally, the Co-free films retain out-of-plane lattice parameters larger than MgO throughout the temperatures shown, supporting Co valence as the primary origin of the temperature-dependent contraction-to-expansion crossover in Figure 2(a). The less abrupt evolution without Cu suggests that Cu coordination contributes to this transition's sharpness.[15] These influences are examined in Sections 3.2 and 3.3. Mg-free CoNiCuZnO further illustrates the interaction between chemistry and kinetics. Films grown at 200 and 500 °C appear single-phase rock salt with opposite strain states, whereas intermediate temperatures produce additional diffraction features.[15] Companion films on sapphire confirm spinel formation at 400 °C (discussed in section 2.3). The proposed interpretation is that low-temperature growth kinetically arrests a rock salt framework containing substantial $Co^{3+}$ before spinel ordering develops; intermediate temperatures permit this ordering, while higher temperatures favor predominantly divalent cations and rock salt retention.[15]

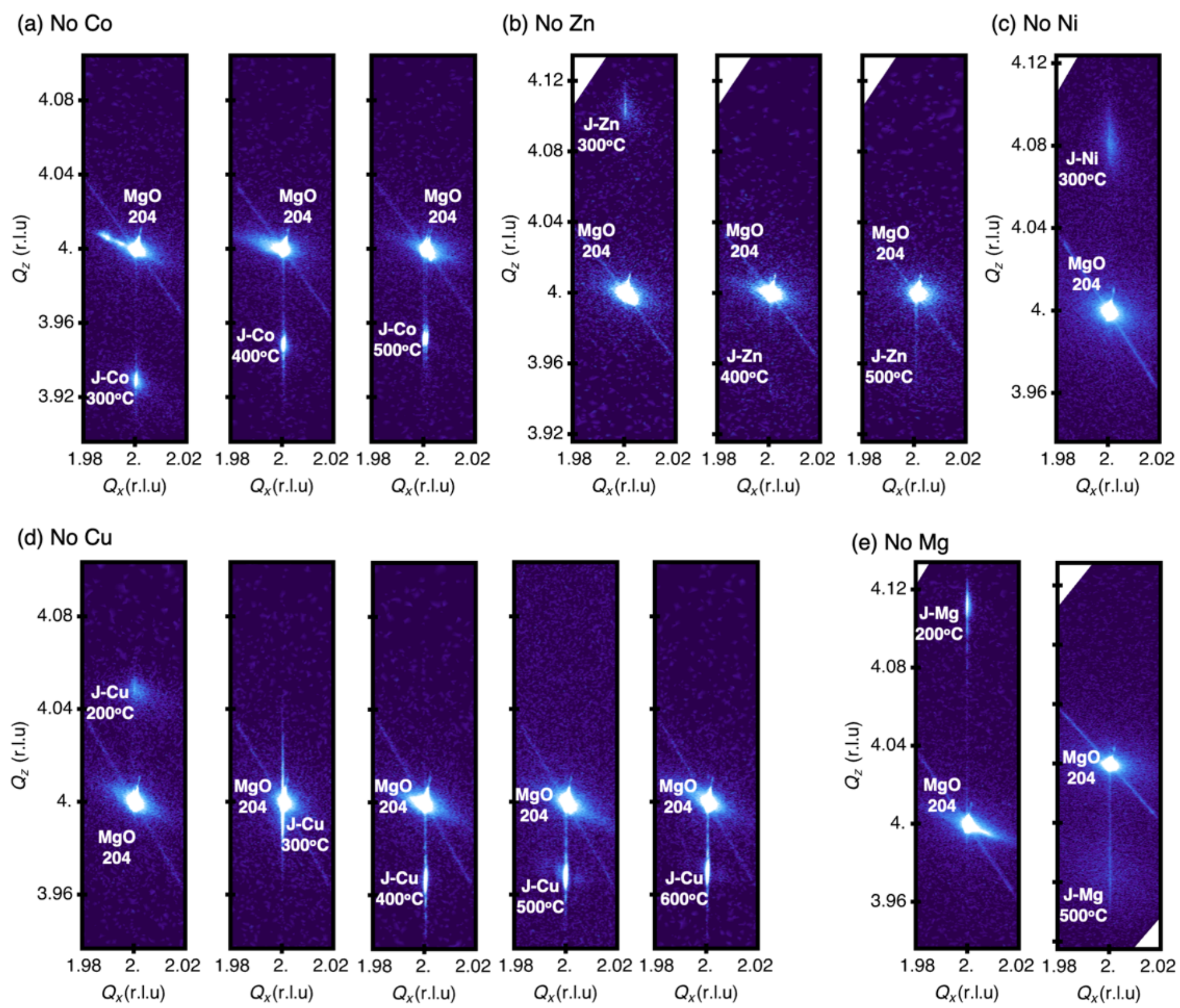

**Figure 3:** A series of reciprocal space maps for four-component derivatives of MgCoNiCuZnO (J14) thin films grown on MgO at varying substrate temperatures (as indicated beneath the film name on each plot), indicating commensurate film growth (shown by MgO 204 reflections) in all cases: (a) MgNiCuZnO (No Co), (b) MgCoNiCuO (no Zn), (c) MgCoCuZnO (no Ni), (d) MgCoNiZnO (no Cu), and (e) CoNiCuZnO (no Mg). Figure 3 is adapted and modified from Almishal, S. S. I. *et al.* Untangling individual cation roles in rock salt high-entropy oxides. *Acta Mater.* **279**, 120289 (2024). [Ref 15]

First-principles decomposition enthalpies complement these observations by considering competing phases with additional oxidation states, unlike mixing enthalpies referenced only to divalent binary oxides. Ni removal gives the least stable derivative and Cu removal the most stable; Co removal reduces the energetic preference for decomposition relative to J14, supporting its destabilizing role under thin-film growth conditions. Calculated decomposition enthalpies correlate with the measured growth-window widths. Together, these results identify Mg and Ni as stabilizers, Cu and Zn as destabilizers, and Co as condition-dependent, demonstrating that cation roles must be interpreted within the synthesis conditions examined.[15] A detailed analysis is provided in Ref.15.

## 3.2 | Co Role in Tuning Lattice Parameter and Nanostructure Formation

Co plays two closely related roles in J14 thin films: its valence tunes the lattice parameter, and its tendency to form spinel enables local ordering and coherent nanostructure evolution. As discussed in Section 3.1, these roles depend on growth conditions. $Co^{2+}$ favors rock salt coordination, whereas oxidation toward $Co^{3+}$ introduces competition with spinel. Substrate temperature and oxygen activity influence this balance, while growth kinetics determine whether the resulting configuration remains dispersed within the rock salt matrix or develops local spinel ordering and discrete nanostructures.[11,12]

The structural response follows from the smaller ionic size of $Co^{3+}$ relative to $Co^{2+}$.[12,27] At fixed oxygen pressure and laser fluence, lower-temperature J14 growth produces a higher net Co valence and a smaller out-of-plane lattice parameter. Higher-temperature growth shifts Co toward $Co^{2+}$ and expands the lattice while the in-plane spacing remains pinned to MgO, as shown in Figure 2(a).[10,12,15] The structural transition accompanying this valence change can be abrupt, producing an approximately 2.6-3.15% out-of-plane expansion range over a temperature interval of less than 100 °C, depending on growth conditions.[10,12,15] Consequently, the films can switch from tensile to compressive epitaxial strain without losing in-plane lattice matching.[10,12,15] Cu coordination also modifies this response[13]; the coupling between the two cations is examined in Section 3.3.

EELS and XAS support the Co valence change without corresponding shifts in Ni or Cu, consistent with the proposed compensation mechanism involving oxygen uptake and cation deficiency to maintain charge neutrality in the ionic rock salt structure.[10,12] An idealized cation-vacancy compensation reaction can be written as

$$2\,\mathrm{Co^{2+}}{}^{\times}_{\mathrm{Co}} + \frac{1}{2}\,\mathrm{O_2}(g) \;\rightleftharpoons\; 2\,\mathrm{Co^{3+}}{}^{\bullet}_{\mathrm{Co}} + \mathrm{V}''_{\mathrm{M}} + \mathrm{O}^{\times}_{\mathrm{O}}$$

in which one vacant cation site compensates every two $Co^{3+}$ ions. $Sc^{3+}$ incorporation can introduce the same compensating defect through

$$\mathrm{Sc_2O_3} \;\longrightarrow\; 2\,\mathrm{Sc^{3+}}{}^{\cdot}_{\mathrm{M}} + \mathrm{V}''_{\mathrm{M}} + 3\,\mathrm{O}^{\times}_{\mathrm{O}}$$

Within this model, the increased cation-vacancy concentration associated with $Sc^{3+}$ incorporation shifts the Co oxidation reaction toward $Co^{2+}$, suppressing $Co^{3+}$ formation.[10] This interpretation is consistent with the lattice-parameter trend in Figure 2(b), where Sc-containing films retain out-of-plane lattice parameters larger than MgO throughout the temperatures examined.[10] The effects of aliovalent additions are discussed further in Section 3.4.

Co also mediates a relaxation pathway through spinel-like ordering formation. Rock salt and spinel share a cubic close-packed oxygen framework but differ in cation-site occupancy. Rock salt fills all octahedral interstices and leaves the tetrahedral interstices empty, whereas spinel occupies half of the octahedral sites and one eighth of the tetrahedral sites.[10,12,15] Its 3:4 cation-to-oxygen ratio requires two thirds of the cations to be trivalent when only divalent and trivalent states are present. In normal-spinel $Co_3O_4$, $Co^{3+}$ occupies octahedral sites and $Co^{2+}$ occupies tetrahedral sites. Evidence of spinel-like local ordering has appeared in several J14-derived systems, particularly those containing Cr, where electron diffraction reveals additional reflections suggestive of spinel ordering.[10,17] However, the clearest atomic-resolution evidence of spinel formation within the rock salt matrix comes from thick J14 films. In films grown at 400 °C, increasing the pulse count from 1200 to 6000 produces coherent spinel nanocuboids at both 1 and 5 Hz.[11] Atomic-resolution imaging resolves the local cation arrangement, while electron diffraction supports spinel symmetry and Co-edge spectroscopy indicates a higher Co valence within the cuboids than in the surrounding matrix, with the Ni and Cu edges unchanged.[11] Their formation is interpreted as a strain-mediated process: the greater stored elastic energy in thicker films favors configurations containing the smaller $Co^{3+}$ ion, and local chemical and structural reorganization reduces the strain energy.[11] The films can therefore retain overall pseudomorphic growth while developing coherent nanoscale heterogeneity.[11]

Therefore, Co sits at the edge of its structural comfort zone, accommodated in rock salt under some growth conditions and drawn toward spinel under others. This sensitivity links growth conditions and defect chemistry to lattice tunability and coherent ordering, with consequences for properties discussed in Section 5.

## 3.3 | Cu Role in Shaping Lattice Parameter and Nanostructure Formation

Cu plays a dual structural role in J14-derived films. Although it destabilizes rock salt relative to competing structures, its Jahn-Teller-active $Cu^{2+}$ ions introduce local, anisotropic distortions that can both generate strain and help accommodate it.[13,14] We propose that these distortions couple with Co valence to modify the lattice response and also influence coherent nanostructure formation.[11,13]

The Cu concentration series in Figure 4 demonstrates this coupling. At 500 °C, increasing Cu content from 0 to 20% produces a nearly linear increase in the out-of-plane lattice parameter. At 300 °C, the same compositional change instead produces a nonlinear contraction, particularly above 5% Cu. The two series have nearly identical lattice parameters without Cu but diverge by approximately 2.6% at the equimolar J14 composition. Cu therefore amplifies the temperature-dependent response discussed in Section 3.2.[13] The proposed interpretation is that, at higher growth temperatures, predominantly divalent Co allows Cu-driven Jahn-Teller elongation to expand the lattice out of plane. At lower temperatures, increasing Cu promotes further Co oxidation, whose contraction outweighs this expansion; local reorientation of the $CuO_6$ distortion axes toward the film plane may also contribute.[13]

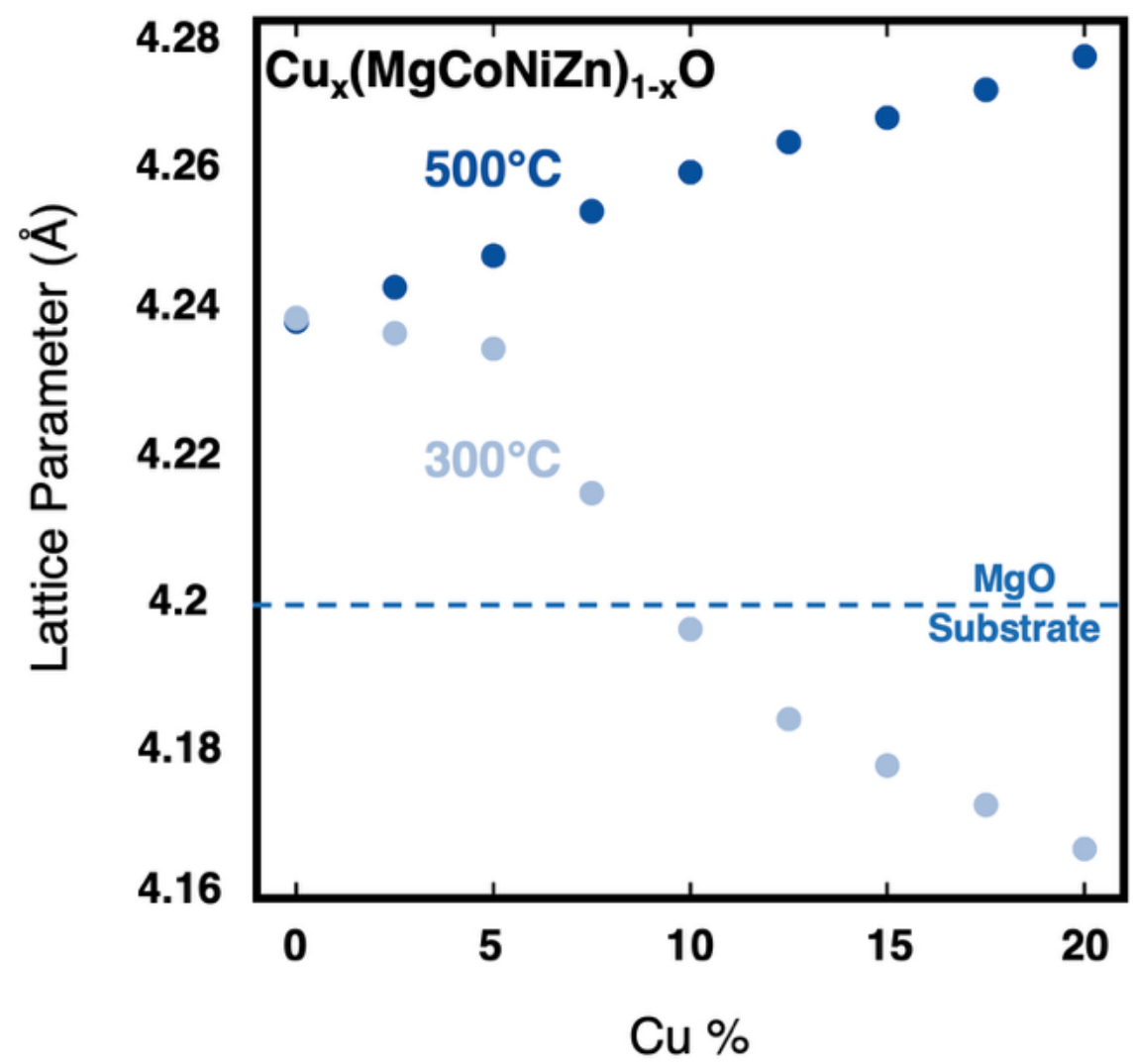

**Figure 4: Contrasting Cu-dependent lattice responses at 300 and 500 °C**: Out-of-plane lattice parameter as a function of Cu concentration for films grown at 300 and 500 °C substrate temperature. The blue dashed line marks the MgO lattice parameter. Figure 4 is adapted and modified from Srikanth, D. *et al.* Jahn-Teller Distortion-Driven Lattice Parameter Tunability in Coherent High-Entropy Oxide Thin Films. Preprint at https://doi.org/10.48550/arXiv.2609.31899 (2026). [Ref 13]

Ca-containing films illustrate how Cu can introduce strain while assisting its accommodation. Removing Cu extends Ca incorporation from approximately 10% to an equimolar 20%, consistent with the large bond-length variations associated with Ca-Cu interactions.[14,20] Nevertheless, at 7.5% Ca, the Cu-containing film has a 3.5% larger out-of-plane lattice parameter than its Cu-free counterpart and retains more pronounced Pendellösung fringes. At 10% Ca, the difference reaches approximately 4.2%; the Cu-containing film remains pseudomorphic to MgO, whereas the Cu-free film relaxes in plane.[14] These observations motivated the proposal that reorientable Jahn-Teller distortions help accommodate in-plane strain while amplifying out-of-plane distortion.[13,14] Cu thus limits Ca incorporation but can improve crystalline quality and preserve in-plane lattice matching within the accessible composition range.[14] Further discussion of Ca incorporation follows in Section 3.4.

Cu also enables a distinct nanoscale ordering pathway. In J14 grown at 400 °C, reducing the pulse repetition rate from 5 to 1 Hz produces coherent CuO-rich nanotweed platelets along {101}.[11,26] The fivefold longer interval between pulses favors their formation, consistent with a diffusion-limited redistribution of Cu within the rock salt matrix.[11,26] Their platelet morphology is attributed to the anisotropic transformation strain associated with Jahn-Teller distortions, supported by microelasticity calculations and phase-field simulations.[26]

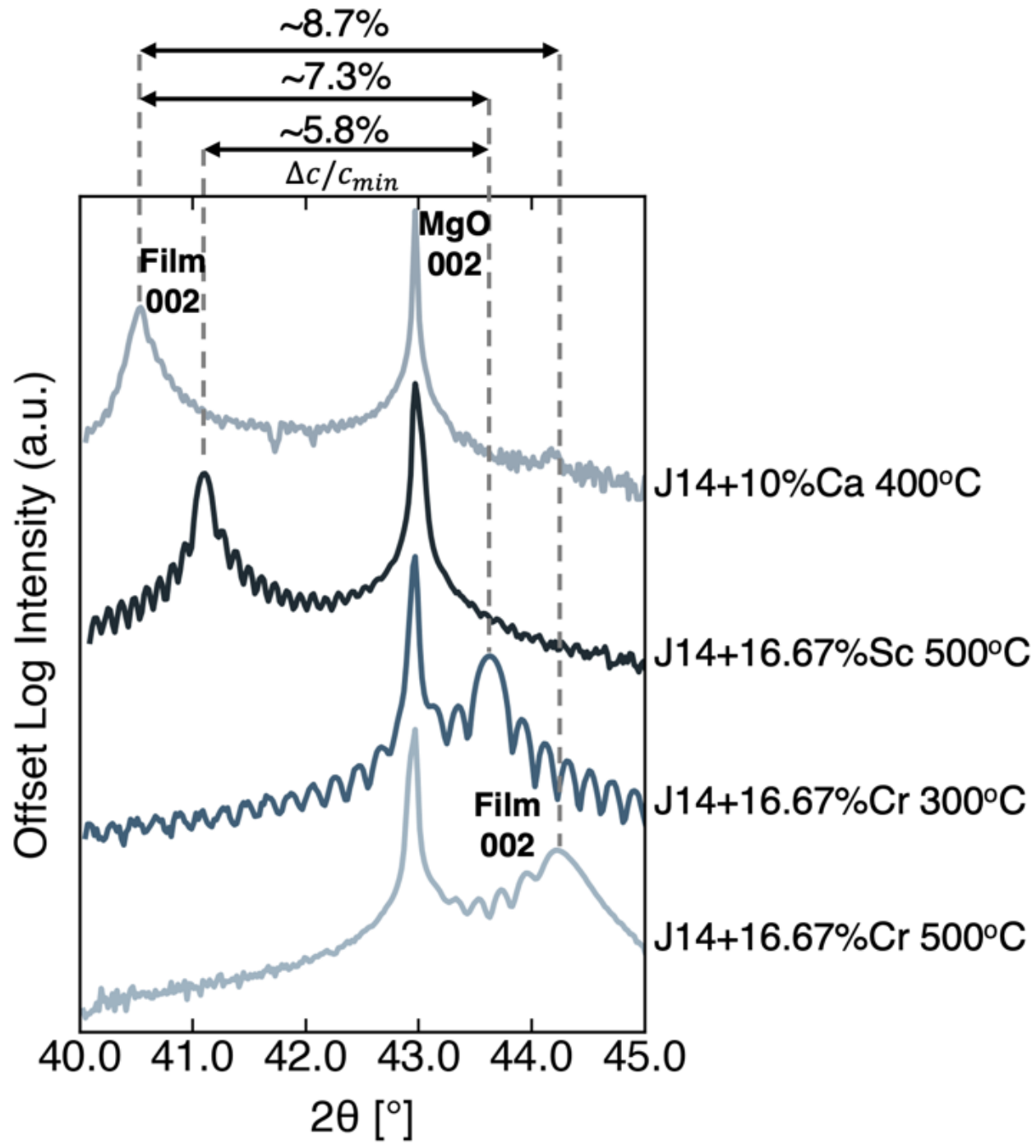

**Figure 5: Expansive lattice parameter tunability from aliovalent and misfit cation addition in MgCoNiCuZnO (J14) epitaxial films.** Sixth-component additions in J14 and selecting favorable growth conditions expands lattice constant tunability considerably, up to 8.7%. $\theta$–$2\theta$ X-ray diffraction scans around the (002) MgO reflection for four rock salt oxide compositions grown on MgO. From top to bottom: J14 + 10% Ca (400 °C), J14 + 16.67% Sc (500 °C), J14 + 16.67% Cr (300 °C), and J14 + 16.67% Cr (500 °C) as indicated to the right of the figure. Figure 5 compiles XRD plots adopted from Almishal, S. S. I. *et al.* Anomalous Crystallinity and Magnetism in Chemically Disordered Coherent Heterostructures. *Adv. Mater.* **38**, e73998 (2026). and Furst, M. *et al.* Maximizing Misfit Cation Solubility in Rock-Salt High-Entropy Oxides. *J. Am. Ceram. Soc.* **109**, e70842 (2026). [ref 10,14]

Thick films grown slowly contain both Cu-rich nanotweeds and the Co-associated spinel nanocuboids discussed in Sections 2.4 and 3.2, while retaining overall pseudomorphic growth.[11,26] Together, Co valence and Cu coordination govern the lattice response and coherent nanoscale ordering, with consequences for properties discussed in Section 5.

## 3.4 | Aliovalent and Misfit Additions as a 6$^{th}$ component

Adding a sixth cation to J14 expands the accessible landscape of structural, valence, and defect configurations. $Sc^{3+}$ and $Cr^{3+}$ probe aliovalent incorporation with contrasting ionic sizes, whereas $Ca^{2+}$ tests substantial size mismatch without changing nominal valence.[10,14] Figure 5 compares their out-of-plane responses, spanning approximately 8.7% between the 10% Ca film grown at 400 °C and JCr grown at 500 °C.[10,14]

$Sc^{3+}$ is only approximately 2.8% larger than the average J14 cation and closely matches high-spin $Co^{2+}$, so its direct size contribution to lattice expansion is small.[10] We posit that its principal influence is instead attributed to defect chemistry.[10] Cation-vacancy compensation of $Sc^{3+}$ is proposed to shift the coupled Co oxidation equilibrium toward $Co^{2+}$, as discussed in Section 3.2. JSc consequently retains an out-of-plane lattice parameter larger than MgO throughout the investigated temperature range, increasing smoothly from 4.32 Å at 300 °C to 4.41 Å at 600 °C while remaining pseudomorphic.[10]

$Cr^{3+}$ introduces a competing influence.[10] Although vacancy compensation may similarly favor $Co^{2+}$, its approximately 15% smaller ionic radius favors contraction.[10] JCr grown at 300 °C has an out-of-plane lattice parameter close to that of J14 at the same temperature but remains contracted at higher temperatures.[10] At 500 °C, additional weak diffraction reflections suggest local spinel-like or doubled-cell ordering, as discussed in Section 3.2.[11,17,28] Sc and Cr therefore enable more than 7% variation in the out-of-plane lattice parameter through the combined effects of ionic size, charge compensation, and local ordering.[10] This range is accessed by changing only the sixth cation, representing 16.67% of the cation composition, and tuning substrate temperature.[10]

$Ca^{2+}$ tests a different incorporation limit. Although isovalent and compatible with rock salt coordination, it is approximately 38% larger than the average J14 cation.[14,20] Bulk synthesis accommodates approximately 4% Ca in Cu-containing J14 and 5% in the Cu-free host.[14] PLD extends these limits to approximately 10% and 20%, respectively, under the appropriate growth conditions; films grown at 400 °C and approximately 1.87 J/cm$^2$ demonstrate how deposition conditions can retain compositions inaccessible through near-equilibrium processing, as discussed in Sections 2.3 and 2.4.[14] Cu introduces a tradeoff within this window. Its removal permits greater Ca incorporation, yet the 10% Ca Cu-containing film remains pseudomorphic while its Cu-free counterpart relaxes in plane.[14] The Cu-containing film also exhibits a 4.2% larger out-of-plane lattice parameter, consistent with the proposed strain-accommodation role of Jahn-Teller distortions discussed in Section 3.3.[14] Thus, size-misfit tolerance depends on the surrounding cations and their local coordination as well as the added ion's radius.[13,14]

Sixth-cation additions therefore expand the accessible configuration landscape of J14-derived films.[10,14] Their ability to accommodate chemical disorder and substantial lattice variation while retaining high crystalline quality and coherence exemplifies the anomalous crystallinity explored in Section 4, alongside its opportunities for heterostructure design. The resulting property consequences are discussed in Section 5.

## 4 | Anomalous Crystallinity in Chemically Disordered Oxide Films

Sections 2 and 3 examined how PLD growth conditions and cation chemistry govern kinetic arrest, lattice tuning, and local ordering in HEO thin films. A recurring feature across these epitaxial systems is the coexistence of extensive chemical disorder with exceptional single-crystalline quality, an intersection we term anomalous crystallinity.[10] This behavior is particularly striking in coherent films, where substantial chemical, valence, and structural variation can be accommodated while preserving lattice registry with the substrate.[10]

This section therefore brings together the preceding observations to examine anomalous crystallinity in HEOs. We posit that the coupled chemical, structural, and valence degrees of freedom, together with rich defect chemistry, provide routes for accommodating local mismatch and compensating chemical or mechanical strain while retaining long-range crystalline order.[10] We consider how this adaptability manifests through extended solubility, thick pseudomorphic growth, crystalline quality relative to simpler compositions, and coherent heterostructure design, before examining related observations beyond rock salts.[10,11,14,15]

## 4.1 | Extended Solubility, Unusual Coordination and Rich Valence Landscape

A first manifestation of anomalous crystallinity is the accommodation of cations with different preferred structures, ionic sizes, and valences within a highly ordered crystal framework, as discussed in Sections 3.1–3.4. J14 and its derivatives illustrate this flexibility: $Zn^{2+}$ occupies octahedral sites despite its tetrahedral coordination in wurtzite ZnO, while $Cu^{2+}$ retains local Jahn-Teller distortions within the rock salt matrix.[12,15,29] Changes in Co valence, coupled with Cu coordination, further modify the lattice while preserving its overall rock salt framework.[10,13] $Sc^{3+}$ and $Cr^{3+}$ introduce charge-compensation requirements, with cation vacancies proposed as a compensating mechanism in the predominantly divalent lattice, whereas $Ca^{2+}$ imposes substantial size mismatch up to 10% incorporation in J14.[10,14] Figure 5 exemplifies the coexistence of these chemical and structural variations with exceptional crystalline quality. Pronounced Pendellösung fringes, together with substrate in-plane lattice matching these films, demonstrate their accommodation within a coherent crystalline framework.[10,14] Local spectroscopy complements these structural measurements by resolving cation-specific valence and coordination environments; interested readers are referred to Ref. 3 for further discussion.

Beyond rock salt, epitaxial $Sr_{0.95}(Ti,Cr,Nb,Mo,W)O_3$, discussed in Section 2.4, illustrates the extended solubility and exhibits a "valence milieu," in which multiple B-site oxidation states coexist within a highly crystalline perovskite framework.[18] This accommodation of local charge diversity while retaining long-range crystalline order is examined further in Section 4.5.

## 4.2 | Coherent Growths to Substantial Thicknesses and under Extensive Pseudostrain

Anomalous crystallinity also manifests in the persistence of pseudomorphic growth to substantial film thicknesses particularly for rock salt oxides grown on MgO. As a coherently strained film thickens, its stored elastic energy per unit area increases, providing a greater driving force for relaxation.[11] The Matthews-Blakeslee model provides a classical framework for estimating the critical thickness for relaxation through misfit dislocations.[30] The persistence of coherence in thick HEO films motivates dedicated studies of how chemical flexibility and growth kinetics influence the critical thickness and relaxation pathways. The following examples illustrate this behavior and provide a basis for such investigations.

Figure 6 illustrates this behavior through reciprocal space maps of a J14 thickness series grown at 450 °C in 50 mTorr $O_2$, with a growth rate of approximately 8 nm/min. The films retain the MgO substrate's in-plane lattice spacing and low mosaicity up to 500 nm. This extended pseudomorphic growth demonstrates that substantial thickness and extensive chemical disorder can coexist with a highly ordered, epitaxially constrained lattice.[31]

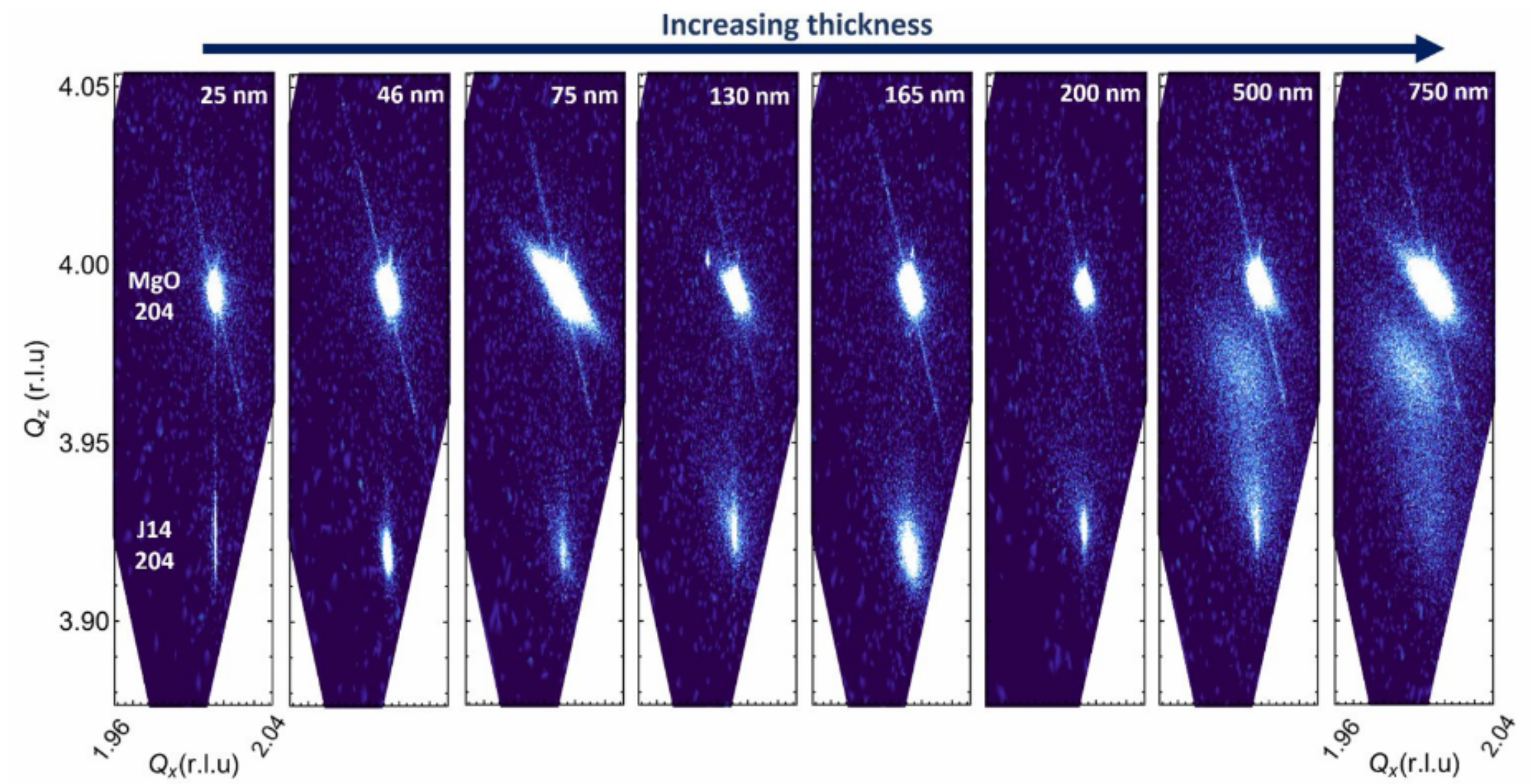

**Figure 6: Coherent MgCoNiCuZnO thin films up to substantial thicknesses:** a series of reciprocal space maps of J14 grown at 450ºC and 50 mTorr $O_2$ with a growth rate of 8nm/min illustrating commensurate film growth and low mosaicity up to 500 nm thickness. Figure adapted from Almishal, S. S. I. Oxides with High Entropy. (The Pennsylvania State University, 2025). [Ref 31]

Coherence also persists across substantial variations in out-of-plane lattice parameter. Following our anomalous crystallinity study in Ref 10, we use *pseudostrain* instead of just *strain* to acknowledge the likely contributions of substrate clamping, cation substitution, and defect chemistry to these lattice responses, with defect chemistry and charge compensation playing a predominant role.[10] Figure 5 shows a pseudostrain range equivalent to approximately 8.7% of the MgO lattice parameter across selected J14-derived films. In particular, the Cu-containing 10% Ca film reaches an out-of-plane lattice parameter of 4.45 Å, corresponding to approximately 5.7% pseudostrain relative to MgO, while retaining substrate in-plane lattice matching.[14] Together, the thickness and composition series highlight the capacity to preserve pseudomorphic growth across substantial thicknesses and large structural variations.[10,14]

The coherent nanostructures discussed in Sections 3.2 and 3.3 provide a complementary perspective on strain accommodation. Cu-rich nanotweeds and spinel nanocuboids develop within films that remain pseudomorphic to MgO.[11,26] In particular, spinel nanocuboid formation in thicker J14 films is interpreted as a strain-mediated process in which local Co valence and structural reorganization reduce the stored elastic energy.[11] These observations suggest that the coupled chemical, structural, and valence degrees of freedom provide additional routes for accommodating strain while preserving crystalline registry.[11]

## 4.3 | Higher Crystalline Fidelity compared with Simpler Compositions

A particularly striking manifestation of anomalous crystallinity, especially in rock salts, is that greater chemical complexity accompanies markedly higher crystalline fidelity under comparable growth conditions. Reducing J14 by one cation produces the five four-component derivatives discussed in Section 3.1. Removing any cation except Cu narrows the single-phase growth window, with Ni removal producing the most restricted window over the temperatures examined.[15] Although Cu removal preserves a broad growth window, it weakens the temperature-dependent out-of-plane lattice response.[15] Thus, reducing the chemical complexity can compromise either the accessible growth window or the structural tunability, depending on which cation is removed.[15]

Further reducing the cation count to three or two reveals an even sharper contrast. Figure 7 compares ternary and binary compositions grown on MgO at 300 and 500 °C under otherwise identical deposition conditions.[10] Most show substantially poorer crystalline quality than the multicomponent HEOs, with some film reflections so weak that high-resolution XRD is needed to resolve them; MgCoCuO additionally exhibits multiphase growth.[10] Selected Mg-containing compositions, including MgNiO and MgCrO, grow well on MgO. For MgNiO at 500 °C, this high quality is expected given its small pseudostrain; its out-of-plane lattice parameter also changes little with temperature.[10] HEOs however push the lattice much further, retaining exceptional crystalline quality under large pseudostrain while enabling thick coherent growth and pronounced temperature-dependent lattice tunability.[10]

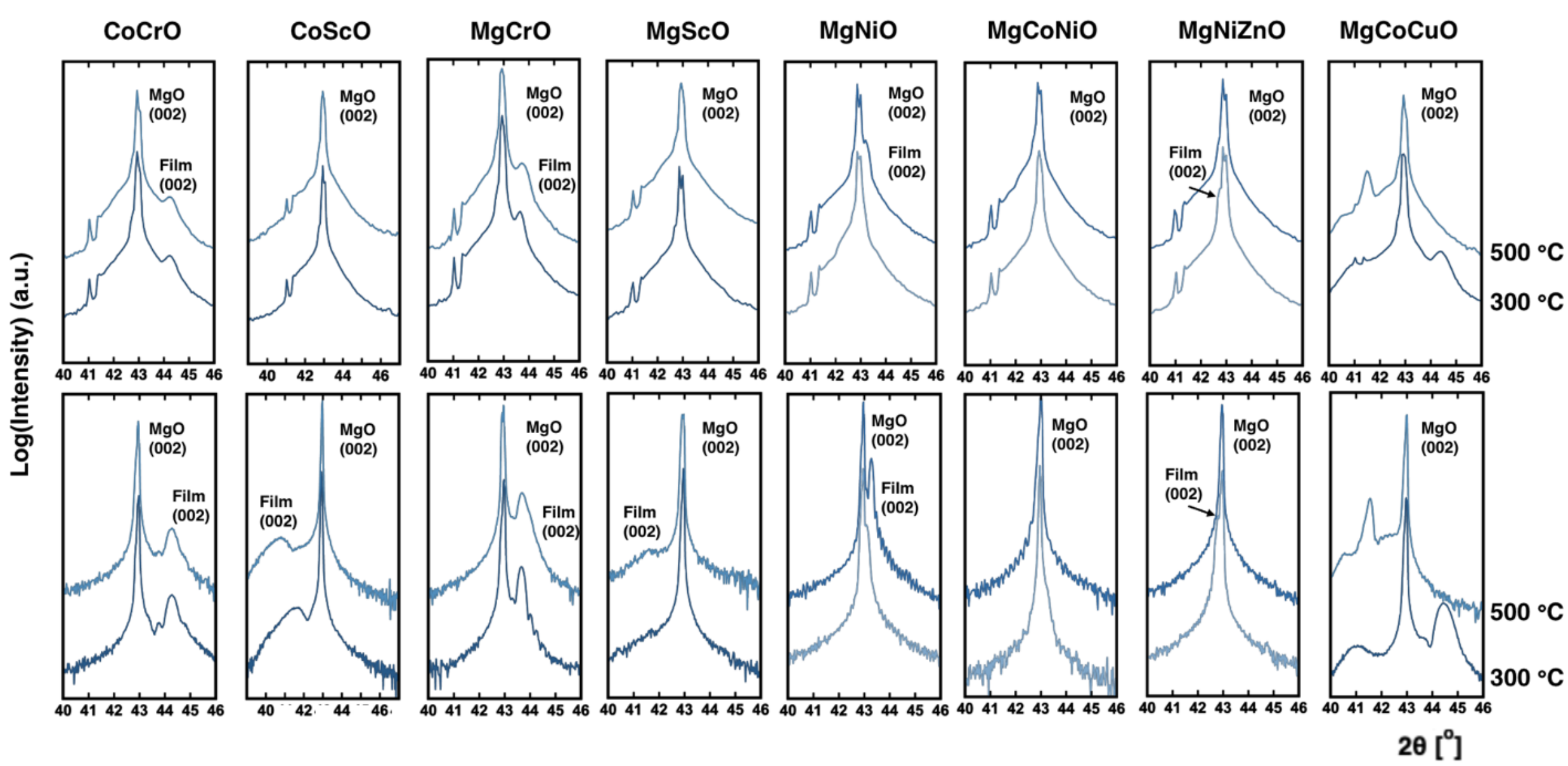

Figure 7. BBHD (Top) and High-resolution X-ray diffraction (bottom) θ–2θ scans around the MgO (002) substrate reflection for a series of binary and ternary oxide thin films (CoCrO, CoScO, MgCrO, MgScO, MgNiO, MgCoNiO, MgNiZnO, and MgCoCuO). For each composition, data are shown for films grown at 300 °C and 500 °C, as indicated.For each composition, data are shown for films grown at 300 °C and 500 °C, as indicated. Figure adapted from Almishal, S. S. I. *et al.* Anomalous Crystallinity and Magnetism in Chemically Disordered Coherent Heterostructures. *Adv. Mater.* **38**, e73998 (2026). [Ref 10]

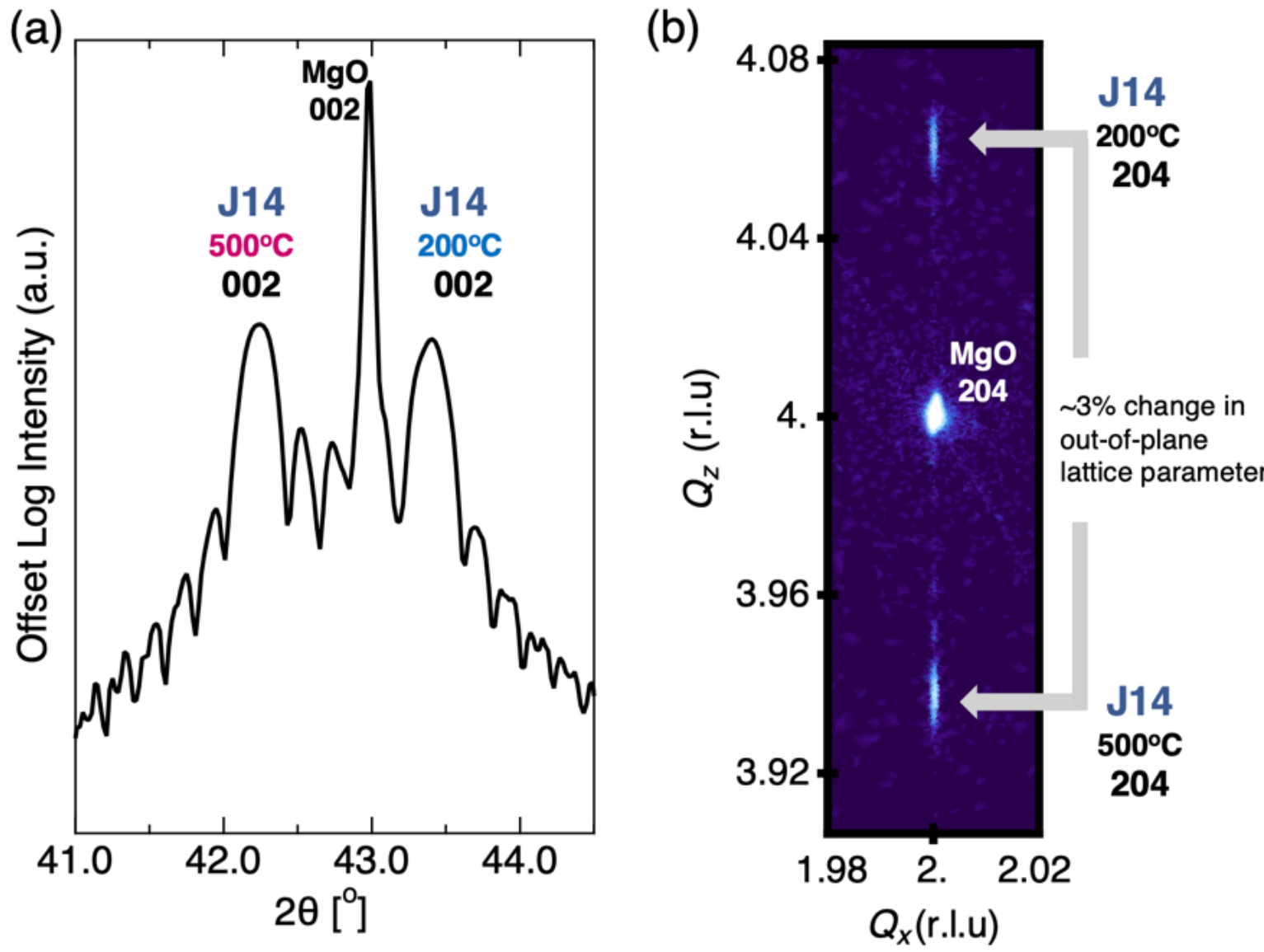

Figure 8: Structural characterization Isochemical Heterostructure Type Z with 3.15% change in out-of-plane lattice parameter: (a) XRD pattern for MgCoNiCuZnO grown at two different temperatures; and (b) the reciprocal space map of the heterostructure. Figure 8 adapted and modified from Almishal, S. S. I. *et al.* Anomalous Crystallinity and Magnetism in Chemically Disordered Coherent Heterostructures. *Adv. Mater.* **38**, e73998 (2026). [Ref 10]

The multicomponent HEO films therefore sustain exceptional anomalous crystalline order across a broader range of thicknesses and lattice configurations. These observations challenge the intuitive expectation that greater chemical complexity should compromise crystalline quality: extensive cation disorder instead coexists with superior long-range crystalline order.[10] Resolving the respective contributions of configurational entropy, cation interactions, defect chemistry, and growth kinetics remains central to understanding the origin of this behavior.

## 4.4 | Lattice Tuning Enabling Heterostructures and Valence Interfaces

The lattice tunability discussed in Sections 4.1 and Section 3.4 enables coherent heterostructures with distinct structural and valence configurations. These interfaces can be created by changing growth temperature within a single composition or by combining different J14-derived compositions, while preserving a common in-plane lattice spacing.[10]

In an isochemical heterostructure, a change in cation valence across the interface creates a *valence interface* while retaining the same nominal composition and crystal framework.[10] Figure 8 illustrates this approach through a J14 bilayer grown at 500 and 200 °C.[10] The layers exhibit an approximately 3% out-of-plane lattice-parameter difference, while reciprocal space mapping shows that both remain in-plane matched to MgO.[10] EELS suggests a higher average Co oxidation state in the lower-temperature layer.[10,12] Their out-of-plane parameters lie on opposite sides of the MgO value, defining a Type Z heterostructure with opposing pseudostrain states.[10] A complementary example is the JSc bilayer grown at 500 and 300 °C. Both layers remain pseudocompressive relative to MgO, defining a Type X heterostructure with an approximately 1.6% out-of-plane contrast.[10] Layer-resolved EELS supports predominantly $Co^{2+}$ in the higher-temperature JSc layer and mixed Co valence in the lower-temperature layer.[10] Both types therefore support valence interfaces with their different pseudostrain configurations.[10]

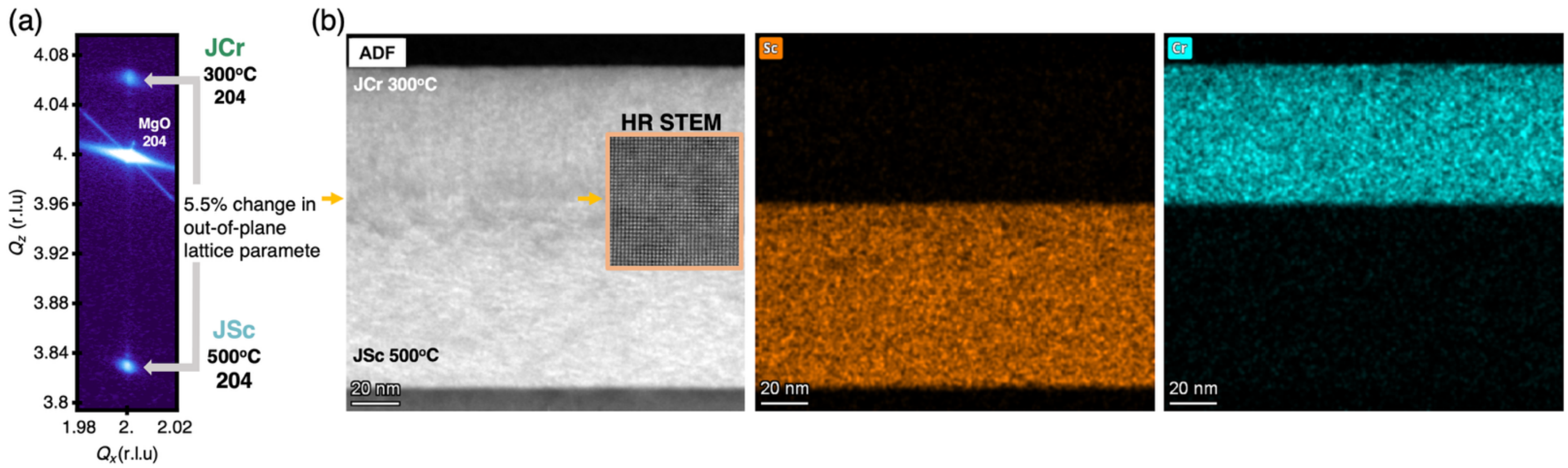

**Figure 9:** (a) Reciprocal-space map around the MgO (204) reflection. The two film peaks lie on the same $Q_x$ value yet separate along $Q_z$, confirming coherent lateral matching and in-plane registry with MgO with different out-of-plane lattice contrast. (b) annular dark-field (ADF) STEM image with corresponding EDS elemental maps across the JSc (500°C)/JCr (300°C) interface, showing an abrupt, chemically sharp boundary without detectable intermixing at the probed length scales. Figure 9 adapted and modified from Almishal, S. S. I. *et al.* Anomalous Crystallinity and Magnetism in Chemically Disordered Coherent Heterostructures. *Adv. Mater.* **38**, e73998 (2026). [Ref 10]

Combining different compositions adds chemical contrast and enables a larger out-of-plane lattice difference while retaining in-plane registry.[10] Figure 9 shows a Type Z heterostructure comprising JSc grown at 500 °C and JCr grown at 300 °C, with an approximately 5.5% out-of-plane lattice-parameter difference. The film reflections share the substrate's in-plane reciprocal coordinate despite their separation out of plane.[10] Atomic-resolution imaging supports lattice continuity across the interface, while Sc and Cr elemental maps reveal an abrupt chemical boundary without detectable intermixing at the probed length scales.[10] Layer-resolved EELS further identifies predominantly $Co^{2+}$ in JSc and mixed $Co^{2+}/Co^{3+}$ in JCr, with an average Co valence near 2.5+ in the latter.[10] The JSc/JCr Type Z heterostructure therefore combines a chemical interface involving a change in one sixth of the cations with a Co valence interface and an approximately 5.5% unit-cell-volume difference, while preserving the in-plane lattice spacing.[10]

These examples show how chemistry and valence can change across an interface while crystalline registry persists, providing a foundation for coherent heterostructure design. Their property consequences are discussed in Section 5. A detailed discussion of all these heterostructures is provided in Ref. 10.

## 4.5 | Beyond Rock Salt: Fluorite, Spinel and Perovskite

The coexistence of extensive chemical disorder with high crystalline quality extends beyond rock salt HEOs. Fluorite, perovskite, and spinel films provide additional examples of how diverse local chemical environments can be accommodated within an epitaxially ordered framework. These systems broaden the manifestations of anomalous crystallinity, although their degree of coherence depends on composition, substrate, and growth conditions.

In fluorite-related HEOs, structural flexibility includes the oxygen-vacancy sublattice. As discussed in Section 2.4, PLD enables $Ce_{0.2}(Y,La,Pr,Sm)_{0.8}O_{2-\delta}$ films to retain a globally fluorite-like structure rather than the bixbyite ordering favored by near-equilibrium processing. This high-symmetry framework accommodates extensive cation disorder together with a substantial oxygen-vacancy population, illustrating that kinetic arrest can preserve disorder on both sublattices.[16] Relatedly, partially relaxed epitaxial $(Hf,Zr,Ce,Gd,Ca)O_2$ films on YSZ combine high crystalline quality with retention of a single fluorite phase even after annealing to 1200 °C.[32]

Spinels introduce additional flexibility through their octahedral and tetrahedral cation sites. Epitaxial $(Co,Cr,Fe,Mn,Ni)_3O_4$ films sustain fully coherent strained growth on $MgAl_2O_4$ and exhibit substrate-dependent magnetic orientation. Coherent rock salt and spinel regions also coexist within these films, paralleling the local structural accommodation observed in J14.[33]

Perovskites expand these possibilities through chemical disorder on the A and B sites and much more diverse cation oxidation states. Single-crystalline $Ba(Zr,Sn,Ti,Hf,Nb)O_3$ films exhibit sharp interfaces and uniform B-site mixing, demonstrating that substantial chemical disorder is compatible with high-quality perovskite epitaxy.[34] Our $Sr_{0.95}(Ti,Cr,Nb,Mo,W)O_3$ films bring together several manifestations of anomalous crystallinity: extended cation solubility, the "valence milieu" introduced in Section 4.1, and high crystalline quality with pronounced Pendellösung fringes and narrow rocking curves sustained to 300 nm thickness, with only modest mosaic broadening as thickness increases. Rare-earth A-site-disordered nickelates provide complementary examples. $(La,Pr,Nd,Sm,Eu)NiO_3$ films exhibit high-quality epitaxial growth, including coherent growth over a range of thicknesses and strain-tunable electronic behavior. Nanoscale chemical inhomogeneity can also coexist with single-crystalline order, demonstrating once again that global crystalline fidelity does not require local chemical uniformity.[35–37] High-quality epitaxial growth, including coherent strain in selected films, also extends to HEOs based on $LaMnO_3$ and related perovskites, including $(Gd,La,Nd,Sm,Sr)MnO_3$,[38] $Nd(Cr,Mn,Fe,Co,Ni)O_3$,[39] and Mn-tuned $La(Cr,Mn,Fe,Co,Ni)O_3$[40]. The latter enables composition-driven magnetic design, discussed further in Section 5.1.

Together, these studies establish that extensive chemical disorder can coexist with exceptional crystalline order across several oxide frameworks.

## 5 | Engineering Functional Properties from Anomalous Crystallinity

Accommodating chemical, structural, and valence variations within highly crystalline films provides routes to functional control through growth conditions, composition, and interface design. While this is not an exhaustive account, we highlight a few examples, beginning with magnetic behavior. In J14 films, growth conditions adjust the $Co^{2+}/Co^{3+}$ balance and influence exchange bias when the HEO is coupled to a ferromagnetic layer.[10,12] Coherent heterostructures extend this control to buried interfaces: the JSc/JCr Type Z (in Figure 9) heterostructure combines chemical and Co valence contrasts while preserving crystalline registry, producing approximately twice the exchange-bias magnitude of constituent single layer.[10] In Mn-tuned $La(Cr,Mn,Fe,Co,Ni)O_3$, compositional control adjusts competing magnetic interactions within a single crystalline layer, enabling antiferromagnetic, ferromagnetic, and frustrated responses.[40] Similarly, epitaxial $(Co,Cr,Fe,Mn,Ni)_3O_4$ films containing coherent rock salt and spinel regions exhibit substrate-dependent magnetic orientation, illustrating magnetic control through strain and local structural accommodation.[33]

Optical and electronic properties provide complementary examples. Lower-temperature J14 films with higher average Co valence exhibit a smaller optical band gap, whereas Cu removal increases the calculated electronic band gap and reduces measured visible absorption.[12,15] Growth conditions and composition therefore provide distinct routes to optical tuning within the retained crystalline framework.[12,15] $Sr_{0.95}(Ti,Cr,Nb,Mo,W)O_3$ thin films combine high electrical conductivity with relatively flat optical transmission despite substantial chemical and valence disorder.[18] Related behavior occurs in highly positionally-ordered high-entropy rare-earth borate single crystals, where weakened absorption features produce enhanced broadband transparency.[41] These examples illustrate how chemical disorder can modify optical absorption while long-range crystalline order persists.

## 6 | Outro

Overall, PLD expands HEO design by coupling compositional freedom with kinetic control over the configurations accessed and retained during growth. The resulting films exemplify anomalous crystallinity, combining extensive chemical disorder with exceptional crystalline quality, lattice tunability, and, in selected systems, coherent interfaces that enable functional engineering. Understanding how configurational entropy, defect chemistry, and growth kinetics sustain this behavior will be essential to translating the observed adaptability into predictive design of new compositions, heterostructures, and properties.

## Acknowledgements

The authors gratefully acknowledge support from NSF MRSEC DMR-2011839. They also thank Matthew Furst, Dhiya Srikanth, Billy Yang, Sai Venkata Gayathri Ayyagari, George Kotsonis,

Yueze Tan, Nasim Alem, John T. Heron, Venkatraman Gopalan, Long-Qing Chen, and Vincent H. Crespi for the insightful discussions that helped shape this work.

## Conflict of Interest

The authors declare no conflicts of interest.